\documentclass[preprint,12pt]{elsarticle}

\usepackage{amssymb}
\usepackage{graphicx}
\usepackage{dcolumn}
\usepackage{bm}
\usepackage{natbib}
\usepackage[T1]{fontenc}
\usepackage[latin1]{inputenc}
\usepackage[english]{babel}
\usepackage{color}
\usepackage{graphics}
\usepackage{pgf}
\usepackage{multimedia}
\usepackage{mathptmx}
\usepackage{helvet}
\usepackage{subfigure}
\usepackage{mathrsfs}
\usepackage{amsfonts}
\usepackage{algorithm,algorithmic}
\usepackage[T1]{fontenc} 
\usepackage{makeidx} 
\usepackage{graphicx} 
\usepackage{dcolumn} 
\usepackage{array} 
\usepackage{amssymb} 
\usepackage{amsmath}
\usepackage{textcomp}
\usepackage{multirow}

\usepackage{subfigure}
\usepackage{eucal}
\usepackage{mathrsfs}
\usepackage[all]{xy}
\usepackage{float} 
\usepackage{amsmath} 
\usepackage{amsfonts} 
\usepackage{bm} 

\journal{Materials Chemistry and Physics}

\begin{document}

\begin{frontmatter}



\title{Magnetocaloric functional properties of Sm$_{0.6}$Sr$_{0.4}$MnO$_3$ manganite due to advanced nanostructured morphology}




\author[1]{V. M. Andrade}
\author[2] {S. S. Pedro}
\author[1] {R. J. Caraballo Vivas}
\author{D. L. Rocco\corref{cor1}\fnref{1}}
\ead{rocco@if.uff.br}
\cortext[cor1]{Corresponding author. Tel.: +55 21 26293834}     
\author[1]{M. S. Reis}
\author[3] {A. P. C. Campos}
\author[4] {A. de A. Coelho}
\author[5] {M. Escote}
\author[5] {A. Zenatti}

\address[1]{Instituto de F\'isica, Universidade Federal Fluminense, 24210-340, Niter\'oi, RJ, Brasil}
\address[2]{Departamento de F\'isica, Universidade do Estado do Rio de Janeiro, 20550-900, Rio de Janeiro, RJ, Brasil}
\address[3]{Divis\~{a}o de Metrologia de Materiais, Instituto Nacional de Metrologia, Qualidade e Tecnologia, 25250-020, Duque de Caxias, Rj, Brazil}
\address[4]{Instituto de F\'{i}sica "Gleb Wataghin", Universidade Estadual de Campinas, Caixa Postal 6165, 13083-859, Campinas, SP, Brazil}
\address[5]{Universidade Federal do ABC, 09210-580, Santo Andr\'e, SP, Brasil}

\begin{abstract}
The magnetocaloric effect (MCE) is the key concept to produce new, advanced, freon-like free, low cost and environmental friendly magnetic refrigerators. Among several potential materials, Sm$_{0.6}$Sr$_{0.4}$MnO$_{3}$ manganite presents one of the highest MCE value in comparison to all other known manganites; however, its studied was only concentrated on the bulk material. To overcame this lack of the information we successfully produced advanced nanostructures, namely nanoparticles and nanotubes of that highlighted manganite by using a sol-gel modified method. High resolution transmission electron microscopy revealed nanoparticle and nanotube diameters of 29 nm and 200 nm, respectively; and, in addition, this technique also showed that the wall of the nanotube is formed by the nanoparticles with 25 nm of diameter. The magnetocaloric potentials, $\Delta S_M$ versus T curves, of the nanostructures were obtained and they are broader than the their bulk counterpart. This increases the useful temperature range of a magnetic refrigerator. But also an undesired M-shape profile for the nanotube sample was observed, due to the rising of a superparamagnetic behavior. These results also evidenced the existence of a nanoparticle size threshold below which the advantage to make the transition wider is no longer valid.
\end{abstract}

\begin{keyword}
sol - gel growth \sep magnetic properties \sep nanostructure

\end{keyword}

\end{frontmatter}


Magnetic refrigeration is a promising technology and therefore attention of researchers to this subject increased in a fast pace\cite{advaenergymaterial_2_1288_2012,naturematerials_13_439_2014}. This technology, based on the magnetocaloric effect, is of easy understanding: application of a magnetic field to a magnetic
material, under adiabatic conditions, makes the temperature of such material to increase. On the other hand, under isothermal conditions
(when the magnetic material is in thermal contact with a thermal reservoir), application of a magnetic field induces a heat exchange
between the material and the reservoir. From these two processes it is possible to create a thermo-magnetic cycle and thus a magnetocaloric
device, like air-conditioners, refrigerators and more. The physical quantities that measure the magnetocaloric potential
are the magnetic entropy change $\Delta S$ and adiabatic temperature
change $\Delta T$\cite{deOliveira201089,PhysRevB.90.104422}. From the applied point of view, a lot of families of materials have been studied by the scientific community\cite{naturematerials_12_52_2013,naturematerials_11_620_2012}, that makes a great effort to find a good material for magnetocaloric application, i.e., a material of low cost, good thermal conductivity, low electrical resistivity, strong magnetocaloric effect (MCE), etc.

Perovskite manganites with general formula R$_{1-x}$A$_{x}$MnO$_3$ (R: trivalent rare-earth and A: bivalent alkaline-earth) is one of these promising families to be used as magnetic coolant material\cite{zb,mr}. In particular, Sm$_{0.6}$Sr$_{0.4}$MnO$_3$ bulk has one of the largest magnetocaloric potential among all of the manganites\cite{jap/113/11/10.1063/1.4795769}, but there is no information about the MC potential of nanostructures of this material.

In what concerns nanostructured materials, interest on these have increased in the last decades due to the emerging new physical and chemical properties. In particular, magnetic properties of those have potential applications on magnetic memory devices, sensors, biology, medicine and catalysts \cite{zhangPhysRevB.58.14167,dormann,apl/98/17/10.1063/1.3584018}. Magnetic materials with particle size at nanometric scale present different magnetic behaviors in comparison to their bulk counterpart; and then superparamagnetism, surface spin glass, large coercivity and large magnetoresistance \cite{dey,pram} are these new emerging phenomena in comparison to the bulk counterpart. These behaviors arise due to the increasing of the ratio between the number of atoms in the surface and the atoms in the core (which usually displays the properties of the bulk material). Thus, as the particle size is reduced, the surface plays a fundamental rule in the magnetic properties, as for instance, reduction of magnetic saturation, producing larger magnetocrystalline anisotropy and different magnetic transition temperature \cite{J}, which make broader the magnetic transition\cite{nosso} in comparison to the bulk sample.

Thus, based on the highlighted points above, the aim of the present work is to produce advanced nanostructured materials of one of the most promising manganite (Sm$_{0.6}$Sr$_{0.4}$MnO$_3$) and optimize its magnetocaloric properties by decreasing the particle size. With smaller particle size the magnetic transition will be broader, and, consequently, the working temperature range of the material will increase.

\section{Experimental techniques}

Sol-gel method (Pechini) was used to prepare nanotube and nanoparticle samples. The main difficulty of this method for nanocrystalline multicomponent oxides is to control of the stoichiometry at nanoscopic level and in this work we could overcame this difficulty. For the synthesis, we used analytical grade samarium nitrate (Sm(NO$_3$)$_3$6H$_2$O), strontium carbonate (SrCO$_3$) and manganese acetate (Mn(CH$_3$COO)$_2$4H$_2$O) weighted accurately; and then these were dissolved into nitric and citric acid solution in de-ionized water. The solution were mixed to obtain a clear solution with molar ratio of Sm:Sr:Mn=0.6:0.4:1. A suitable amount of polyethylene glycol was added to the solution as polymerizing agent. In order to evaporate the excess of solvents and to promote polymerization, the solution was submitted to 343 K for 6 h and then a yellow transparent viscous solution was obtained. A part of the solution was heated in a furnace for 8h at 973 K and a final black Sm$_{0.6}$Sr$_{0.4}$MnO$_3$ powder was obtained. The deposition of nanotubes were made in ordered porous alumina with 200 nm of diameter for 2 h in vacuum (pore wetting technique) and treated for 8 h at 973 K.  The bulk sample were prepared by conventional solid-state reaction from stoichiometric amounts of Sm$_2$O$_3$; SrCO$_3$, and Mn$_2$O$_3$. The powders were ground,  mixed and then calcined in air at 1373 K for 24 h. The resulting powder was reground, pressed into pellets, and then sintered at 1623 K during 36 h.

X-ray powder diffraction data were obtained at room temperature, using a Bruker AXS D8 Advance diffractometer with Cu-K$\alpha$ radiation ($\lambda$ = 1.54056 \AA{}) for nanostructures, and with Fe-K$\alpha$ ($\lambda$ = 1.936087 \AA{}) for bulk sample. Data were collected in the 15$^{o}$ < 2$\theta$ < 85$^{o}$ range in a Bragg-Brentano geometry, with a step size of 0.02$^{o}$ and a counting time of 0.1 s per step. To confirm the formation of the nanotubes, high resolution transmission electron microscopy (HRTEM) technique was employed. For these analysis the samples were diluted in alcohol. Magnetic measurements was carried out using a commercial Superconducting Quantum Interference Device (SQUID).

\section{Crystal structure and morphology}

\begin{figure}
\begin{center}
\includegraphics[width=9cm,angle=-90]{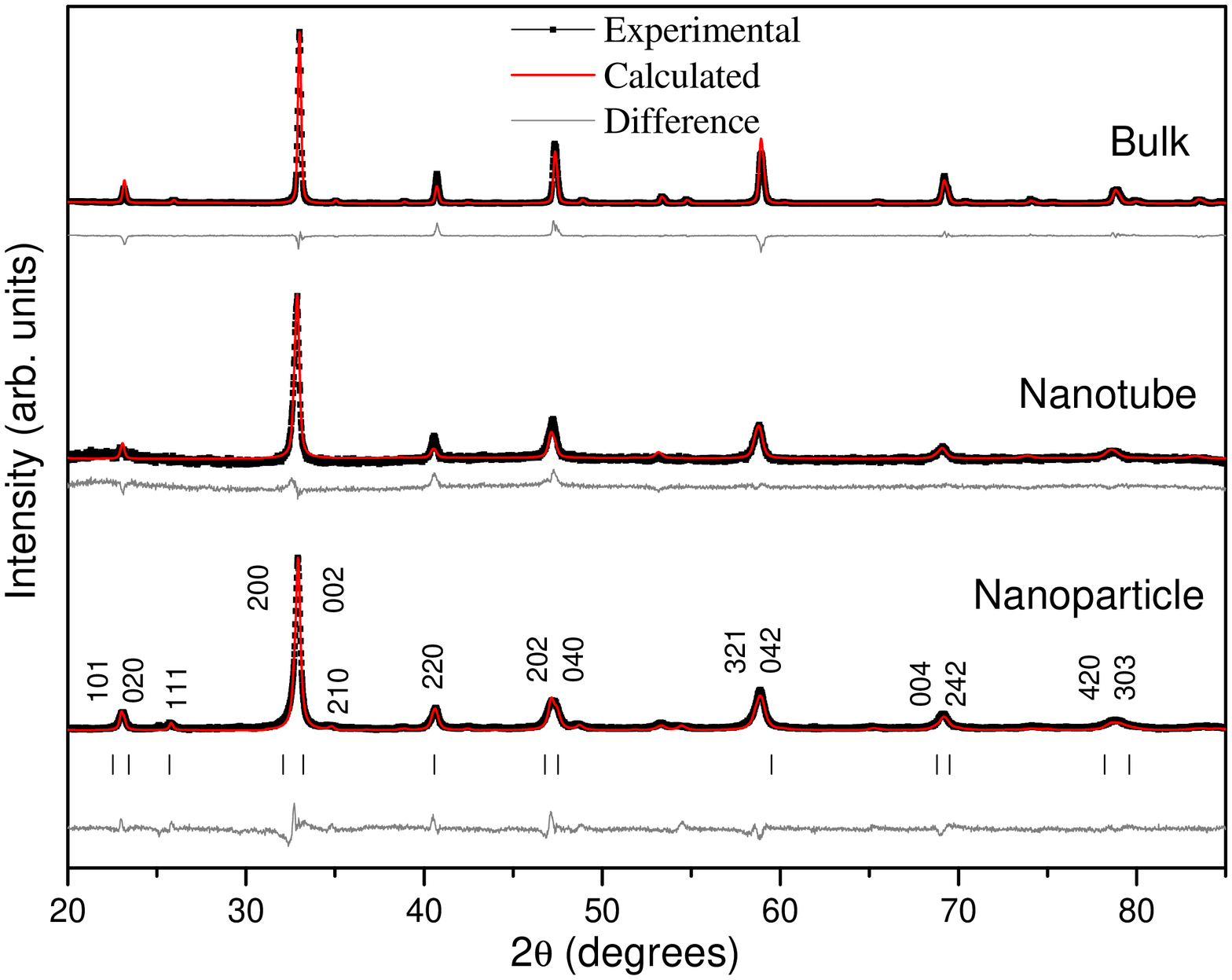}
\end{center}
\caption{Powder diffractograms for the Sm$_{0.6}$Sr$_{0.4}$MnO$_{3}$ nanoparticles, nanotubes and bulk samples.\label{DRXTODOS}}
\end{figure}

X-ray powder diffraction of the nanostructures and bulk are shown in the Fig.\ref{DRXTODOS}; the results confirm the formation of pure Sm$_{0.6}$Sr$_{0.4}$MnO$_3$ crystalline phase (space group \textit{Pnma} ortorhombic system). Those data were refined by the Rietveld method (bottom of Fig. \ref{DRXTODOS}) and the crystallographic parameters and reliability factors obtained are into Table \ref{crystaldata} and are in good agreement with previous results \cite{jap/113/11/10.1063/1.4795769}. Using the Scherrer equation\cite{nosso} was possible to estimate the average particle diameter $D$ from the nanoparticle and nanotube (note the nanotube wall is composed of nanoparticles) which are 29 nm and 15 nm, respectively (Table \ref{crystaldata}).

\begin{table}
\begin{center}
\caption{Refined crystallographic data and reliability factors for Sm$_{0.6}$Sr$_{0.4}$MnO$_{3}$ nanoparticle, nanotube and bulk samples. The nanoparticle size $D$ obtained by the X-ray and HRTEM are also presented.\label{crystaldata}}
    \begin{tabular}{|c|c|c|c|c}
     \multicolumn{2}{l}{Parameter} & \multicolumn{2}{l}{Samples}\\\hline\hline
                        & Nanoparticle    & Nanotube     & Bulk         \\\hline\hline
    $a$ (\AA{})         & 5.4494(6)        & 5.4475(9)       & 5.4448(8)        \\
    $b$ (\AA{})         & 7.6429(8)        & 7.6554(8)       & 7.6729(7)        \\
    $c$ (\AA{})         & 5.4249(6)        & 5.4244(8)      & 5.4420(7)       \\
    $V$ (\AA$^3${})         & 225.84(2)        & 226.61(8)      & 227.35(1)       \\
    $\alpha$ (deg)      & 90            & 90              & 90            \\
    $\beta$ (deg)       & 90            & 90              & 90            \\
    $\gamma$ (deg)      & 90            & 90              & 90            \\
    R$_{p}$ (\%)        & 19.6          & 20.1            & 14.5          \\
    R$_{wp}$ (\%)       & 13.1          & 25.2            & 18.7          \\
    R$_{exp}$ (\%)      & 7.41          & 18.9            & 13.4          \\\hline
    At. position     & (x,y,z)       &(x,y,z)           &(x,y,z)         \\
    Sm and Sr        & (0.02773, 1/4, 0.0039)       &(0.0010, 1/4, 0.0030)           &(0.0272, 1/4, 0.0050)         \\
    Mn               & (0, 0, 1/2)                  &(0, 0, 1/2)                     &(0, 0, 1/2)         \\
    O1               & (-0.02096, 1/4, 0.5644)      &(0.0030, 1/4, 0.4771)           &(0.0010, 1/4, 0.4924)         \\
    O2               & (0.24994, 0.48939, 0.3133)   &(0.4793, -0.0405, 0.1879)       &(0.2439, 0.6107, 0.2427)         \\
     $\chi$          & 1.76          & 1.33             & 1.39           \\\hline\hline
    $D_{X-ray} (nm)$ & $29 \pm 7$    &$15 \pm 4$        & -            \\\hline
     $D_{HRTEM}(nm)$ & $45  \pm 8$    & $25 \pm 4$        & -            \\\hline\hline
            \end{tabular}
\end{center}
\end{table}

\begin{figure}
\center
\includegraphics[width=5cm,angle=-90]{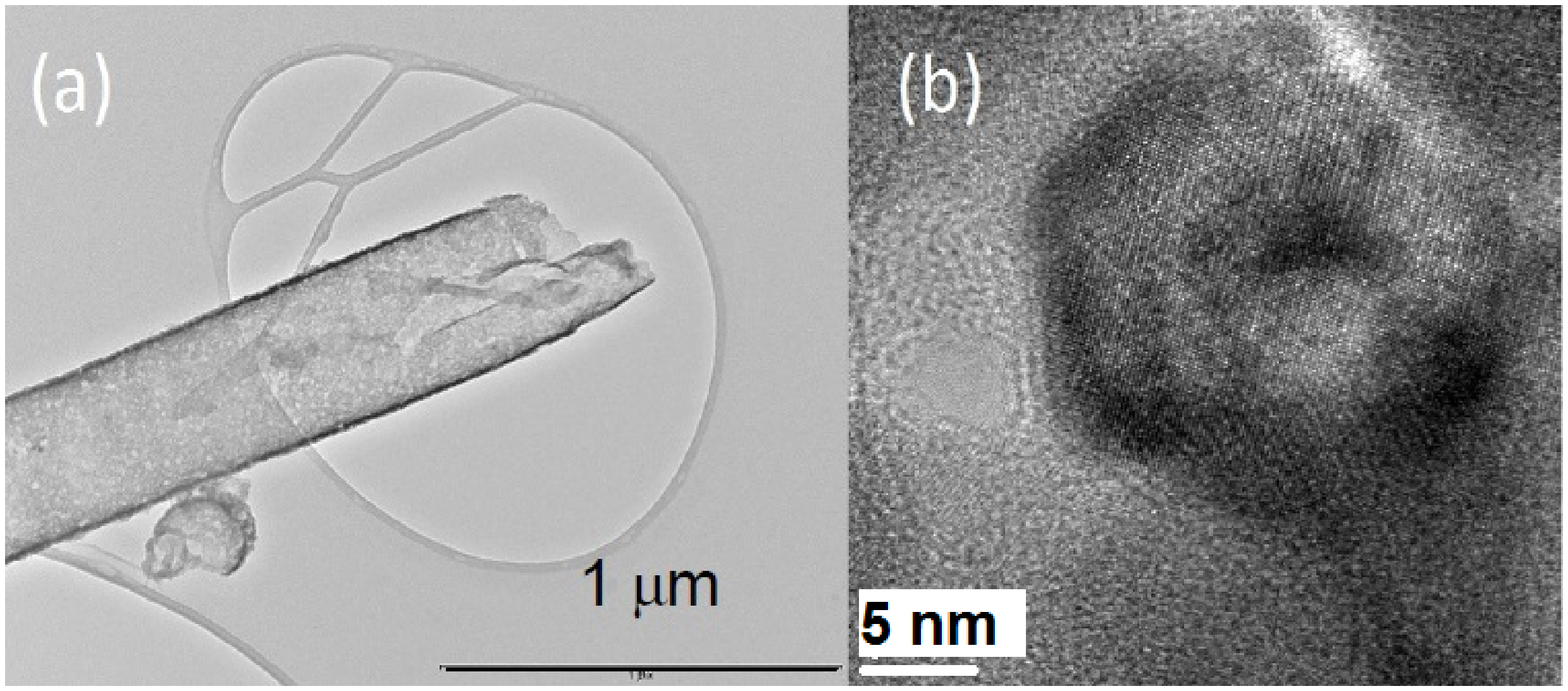}
\caption{Transmission Electron Microscospy for (a) nanotube and (b) high resolution mode used to verify the features of particle that constitutes the wall of the nanotube. \label{TEM}}
\end{figure}

In order to confirm the formation of the nanotubes, HRTEM image were done. In figure \ref{TEM}(a) it is possible to see the nanotube diameter of c.a. 200 nm, while figure \ref{TEM}(b) shows the image of a single nanoparticle that composes the nanotube wall. Note the nanotube is formed by an assembly of nanoparticles with average diameter of $\sim25$ nm. This value is in good agreement with the obtained value by XRD refinement data ($\sim15\pm4$ nm - see Table \ref{crystaldata}).

\section{Magnetocaloric potential evaluation}

\begin{figure}[h]
\center
\includegraphics[width=10cm,angle=-90]{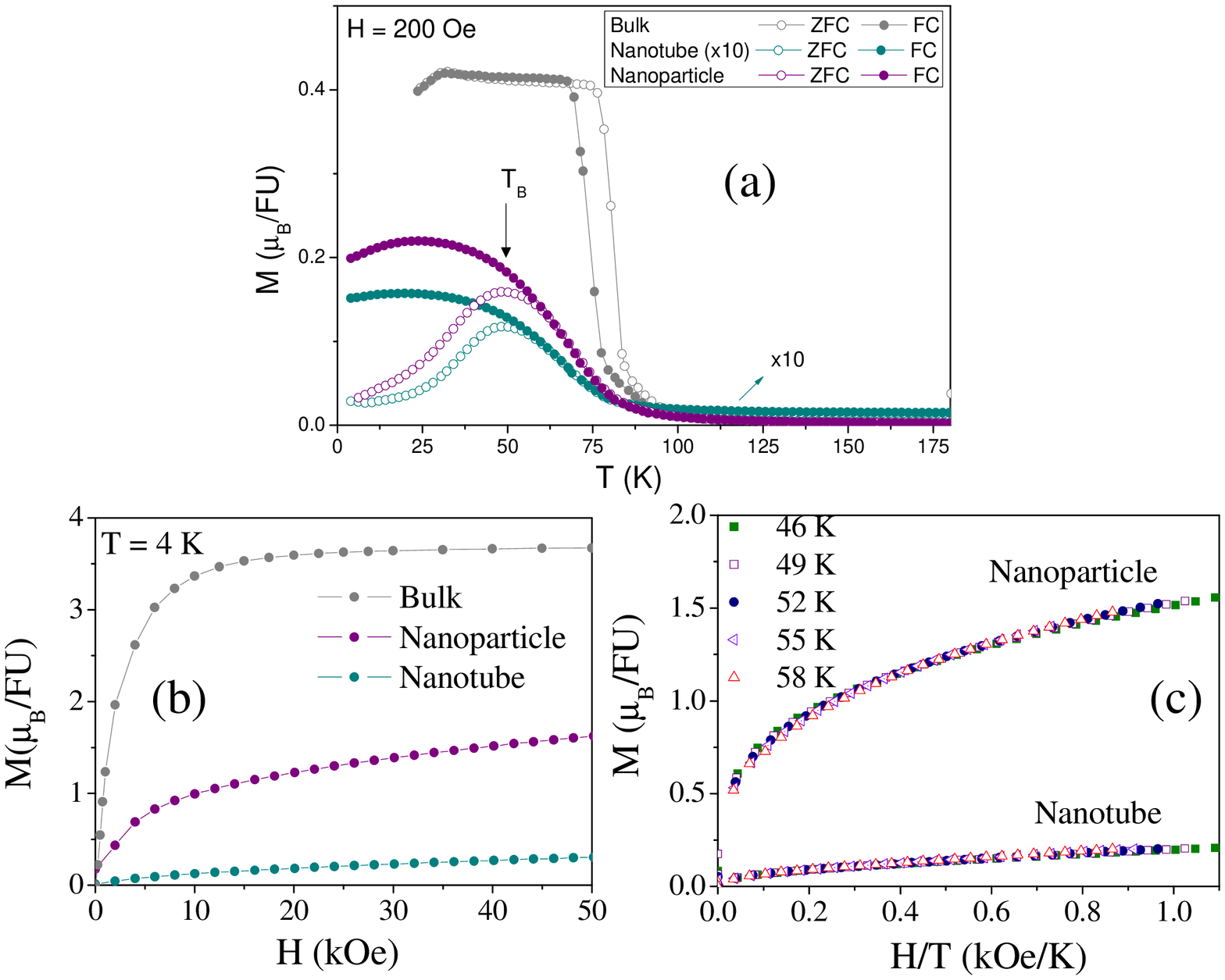}
\caption{(a) Temperature dependence of field-cooled (FC) and zero field-cooled (ZFC) magnetization for nanoparticle, nanotube and bulk of Sm$_{0.6}$Sr$_{0.4}$MnO$_{3}$ manganite. Nanotube magnetization was multiplied by a factor 10 for better visualization. Bottom-(b): magnetization as a function of external magnetic field, at 4 K. Bottom-(c): magnetization as a function of $H/T$ presenting an evidence of superparamagnetic behavior for nanotubes and nanoparticle. \label{MxTtodos}}
\end{figure}

Due to the different morphology of the prepared samples, as detailed before, it is highly expected a different magnetic behavior for those samples; and indeed it was found, as shown on figure \ref{MxTtodos}-(a), that presents the magnetization curves as a function of temperature for all samples. Data were collected in zero field cooled (ZFC) and field cooled (FC) regimes. It is possible to observe that the bulk sample shows a thermal hysteresis, as expected and observed in other works\cite{jap/113/11/10.1063/1.4795769}, in which is a signature of first-order magnetic transition. On the other hand, magnetization of nanoparticles and nanotubes, without thermal hysteresis close to $T_c$, presents a peak around 50 K, which is characteristic of superparamagnetic (SPM) systems, and, in addition, the curve is broader than that one of the bulk. This result indicates that the magnetocaloric potential curve ($\Delta$ S) for nanostructures will be wider.

Moreover, figure \ref{MxTtodos}-(a) shows that the magnetization value at low temperature is very different for all samples, which can be better understood analyzing the magnetization as a function of applied magnetic field at 4 K (see figure \ref{MxTtodos}-(b)). We can see that the nanostructures do not saturate completely in fields up to 50 kOe, but using the M \textit{vs} 1/H curves (not shown), it was possible to obtain the saturation magnetization of 0.4 $\mu_\text{B}\text{/FU}$, 2 $\mu_\text{B}\text{/FU}$ and 3.6 $\mu_\text{B}\text{/FU}$ for the nanotube, nanoparticle and bulk, respectively. Note the magnetization in high magnetic field decreases by decreasing the size of the nanostructure due to the increasing of the surface/volume ratio; and this behavior was already noted in other works\cite{nosso}. The magnetic state of the nanoparticle surface is different from the core (volume), and in fact it is still a point of discussion: some researchers claim that manganite nanoparticles have a surface with ferrimagnetic order \cite{nosso}; other groups focus on the magnetically dead surface \cite{JAP_63922_2012} and some other works argue that the surface presents only a canting effects \cite{APL1}. In spite of this divergence about the magnetic arrangement on the nanoparticle surface, all the works agree that the surface plays a fundamental role in the saturation magnetization.

In addition, in what concerns the superparamagnetic evidences in the nanostructured sample, we used a simple criterion for analyzing such behavior: according to Bean and Livingston \cite{bean}, a system can be considered superparamagnetic if $M$ \textit{vs.} $H/T$ for several temperatures overlap around $T_B$; and indeed this occurs, as can be seen in figure \ref{MxTtodos}-(c). Other works indeed agree with this assumption\cite{PhysRevB.45.9778}.
\begin{figure}[h]
\center
\includegraphics[width=6cm]{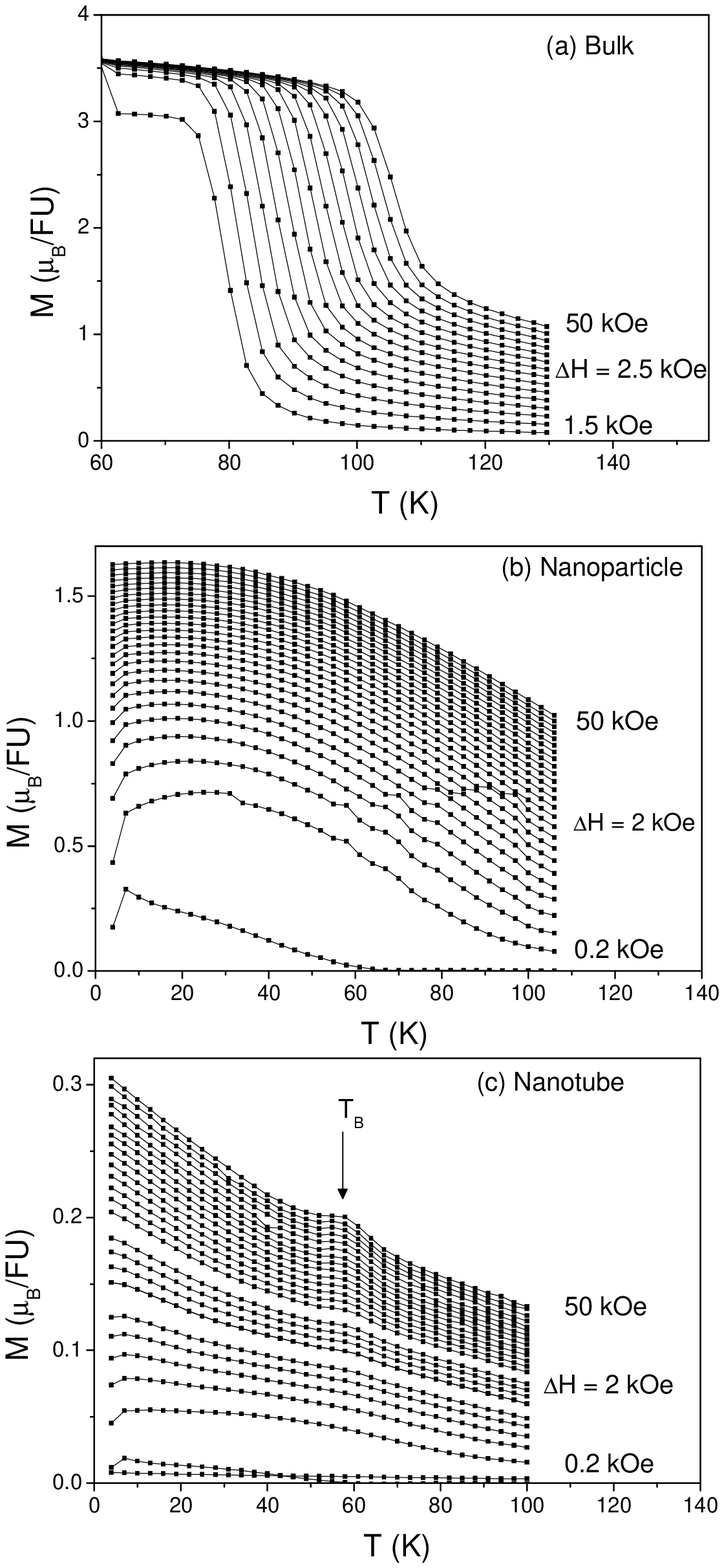}
\caption{Magnetization as a function of temperature for several values of external magnetic field for (a) bulk, (b) nanoparticle and (c) nanotube.\label{MxTvsH}}
\end{figure}

\begin{figure}[h]
\center
\includegraphics[width=5.8cm]{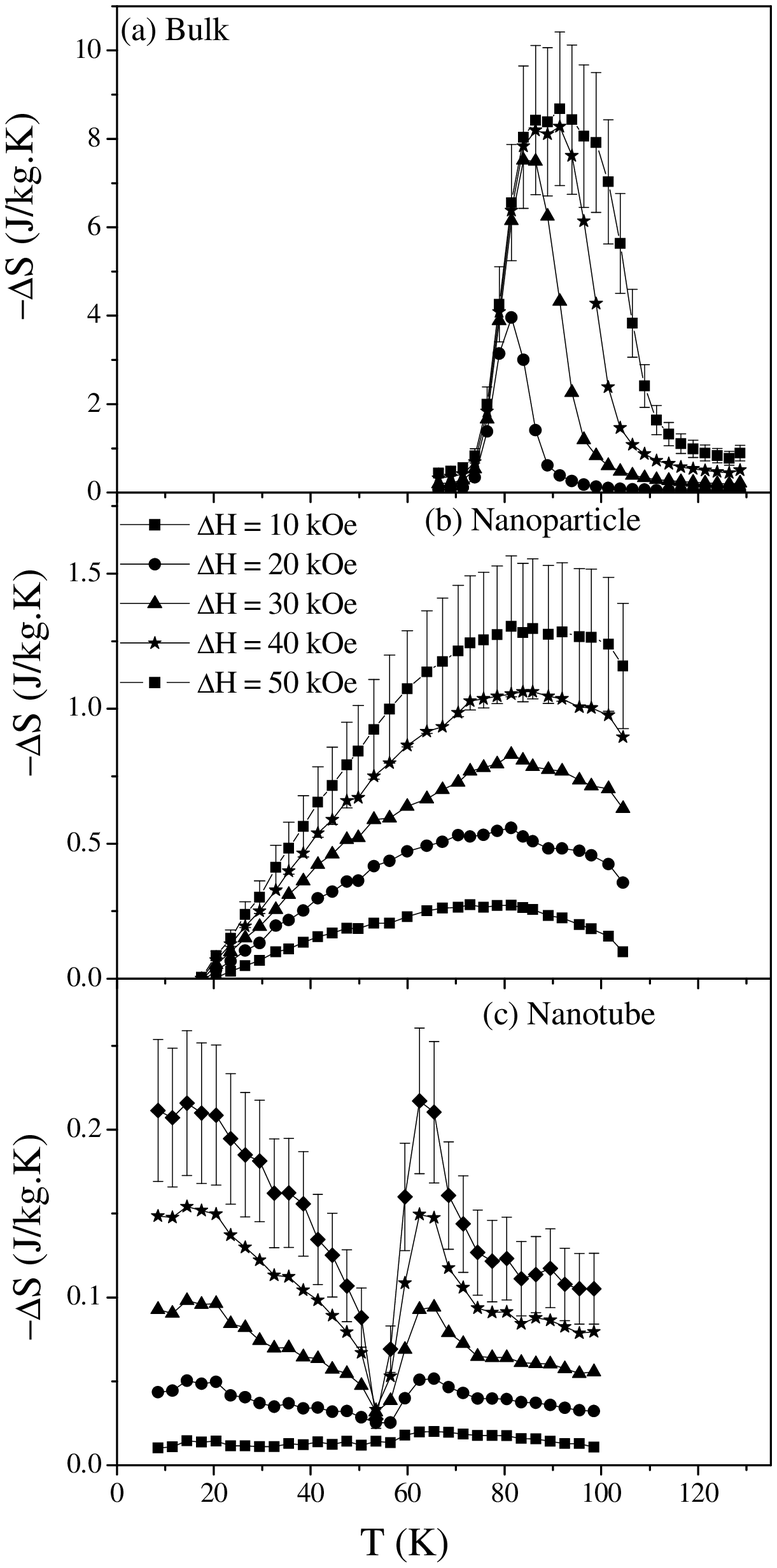}
\caption{Entropy change as a function of temperature for some values of external magnetic field for (a) bulk, (b) nanoparticle and (c) nanotube. The error bars were only presented for 5 T curve for clarity purpose.\label{emc}}
\end{figure}
\begin{figure}[h]
\center
\includegraphics[width=5.6cm]{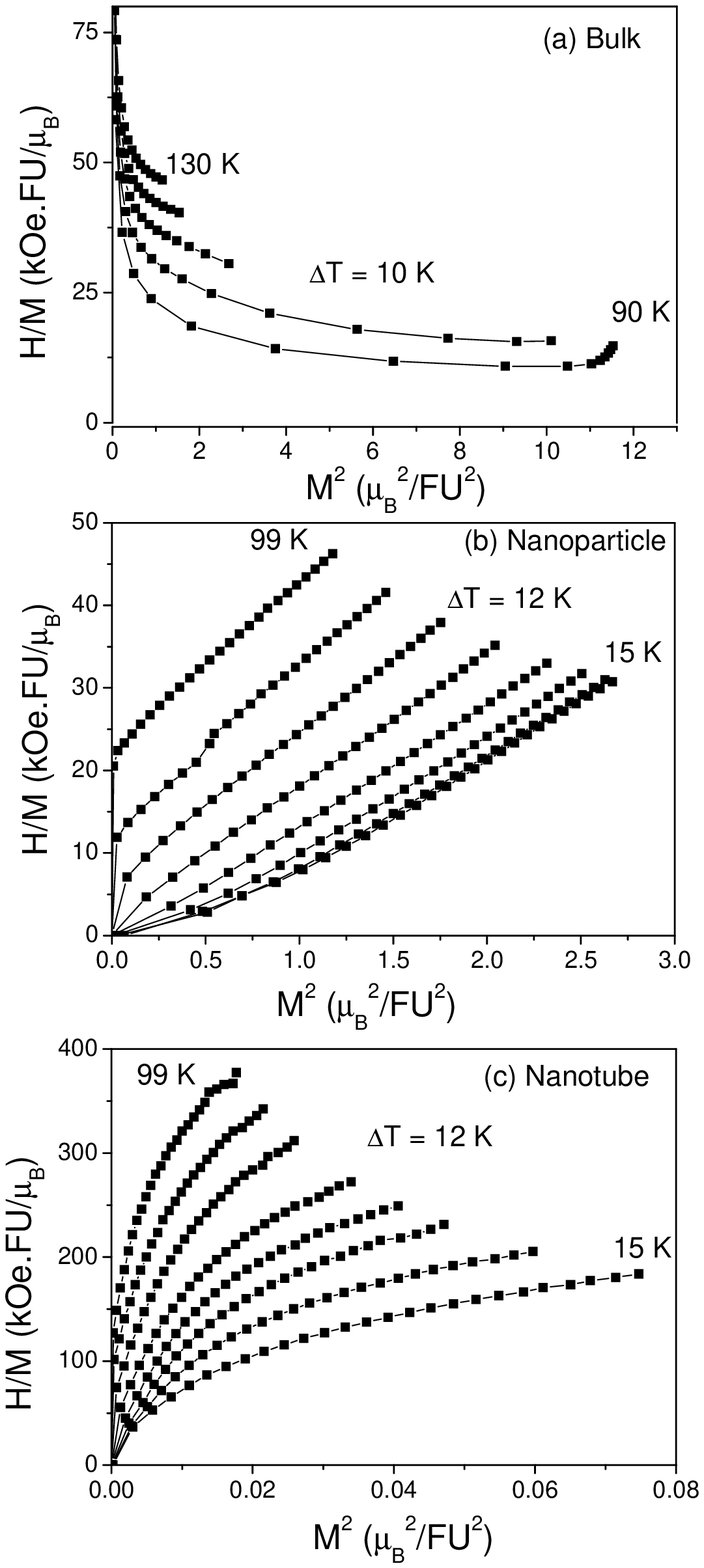}
\caption{Arrott plots for different temperatures for all samples analyzed.\label{arrotplot}}
\end{figure}
Magnetization as a function of temperature for different applied magnetic fields are shown in figure \ref{MxTvsH} for (a) bulk, (b) nanoparticles and (c) nanotubes. The magnetic entropy change was determined from the isothermals (M versus H curves transposed of the M(T) presented in figure \ref{MxTvsH}) using the integral version of a Maxwell relation\cite{PRL_78_1997_4494}:
\begin{equation}
\Delta S=\int_0^{H_f}\frac{\partial M(T,H)}{\partial T} dH
\end{equation}
The obtained magnetic entropy change are shown in figure \ref{emc} for (a) bulk, (b) nanoparticles and (c) nanotubes, for $\Delta$H = 10, 20, 30, 40 and 50 kOe.  Note indeed one of the goal of this work was reached, since we could make broader the magnetic entropy change of this highlighted Sm$_{0.6}$Sr$_{0.4}$MnO$_{3}$ manganite by producing nanoparticles. In other words, this quantity changes from a sharp A-shape, for the bulk sample, to a much more smoothing curve, for the nanoparticles. In what concerns the nanotubes, note a undesired M-shape profile was found and it is directly related to the nanoparticle size (25 nm) that composes the nanotube wall (in contrast to the 29 nm particle size of the pure nanoparticle sample). We mean, it depends on the nanoparticle size because for this case (nanotube), superparamagnetic behavior is more pronounced (note the valley occurs at the blocking temperature at 50 K).

Other important issue is the character of the magnetic transition. As can be seen in figure \ref{arrotplot}, which shows the Arrott Plot, the bulk curves present a negative slope ($B$ parameter of the Landau expansion\cite{livromario}), for low values of magnetization, which indicates a first-order magnetic transition according to Banerjee's criterion \cite{banerjee}; while the two nanostructured samples present a positive $B$ parameter, i.e., transition of second-order character according to same criterion. The bulk suffers a first-order magnetic transition at $T_C$ coupled to a volume change ($\Delta V/V$) of 0.1 \% \cite{abramovich}, which becomes second order as particle decreases. We should remember that the nanoparticle is composed of two different parts; the inner is a core where double exchange interaction dominates and promotes ferromagnetic behavior. The outer part is a layer where magnetic interactions are clearly modified by defects, vacancies, stress, and broke bonds directing to a disordered magnetic state. Hueso et al. \cite{Huesol/jap/91/12/10.1063/1.1476972}  argues that the center part always retains the intrinsic first-order magnetic transition of the bulk compound, while the disordered outer layer is more likely to undergo a second-order transition, from the disordered state into the paramagnetic. The composition of both transition hides the presence of the first-order one. In this way, the overall result in the smallest particles is a second-order transition, although both contribution should be present at the same time.

Thus, the aim of this paper was reached by producing nanostructure of the Sm$_{0.6}$Sr$_{0.4}$MnO$_{3}$ by using a sol-gel modified method. In addition, the magnetic transition of the nanostructure was smoothed, and, thus, the magnetic entropy change curve is wider, instead of a sharp A-shape of the bulk sample. However, a drawback arose: the magnitude of the effect, that decreases from 8.2 J/kg.K for the bulk sample down to 1.2 J/kg.K for the nanoparticles (at 50 kOe of magnetic field change). This decreasing is associated to the decreasing of magnetic saturation, as discussed before (see figure \ref{MxTtodos}-bottom-left). Indeed, it is not as bad as it seems, because the important quantity for magnetocaloric applications is the relative cooling power (RCP); i.e., the maximum value of magnetic entropy change times the full width at half maximum. Thus, the RCP for the bulk sample is 246 J/kg, while it is 132 J/kg for the nanoparticle; and are therefore comparable. Some final words: the decreasing of the maximum magnetic entropy change for the nanoparticles is associated to the decreasing of magnetic saturation, as discussed before (see figure \ref{MxTtodos}-bottom-left).

\section{Conclusions}

Bulk, nanoparticles and nanotubes of Sm$_{0.6}$Sr$_{0.4}$MnO$_3$ manganites were successfully produced by the modified sol-gel method. The structural properties were investigated by the DRX and HRTEM techniques, which showed that nanoparticles and nanotubes present average diameter of 29 and 200 nm, respectively. In addition, we also could determine that nanotube wall is composed of nanoparticles with average diameter of 25 nm. Magnetization measurements reveled a possible superparamagnetic behavior of these nanostrucutres due to the merged M \emph{vs} H/T curves. In addition, in this work was possible to make broader the magnetic entropy change curve of the highlighted manganite. More precisely, this quantity changes from a sharp A-shape (bulk) to a much broader curve (nanoparticle), covering c.a. 110 K (full width at half maximum). However, for nanostructures the maximum of magnetic entropy change is significantly lower than the bulk due to suppression of first-order magnetic transition by nanostructuring. On the other hand, nanotubes, constituted of smaller nanoparticles (25 nm), have the superparamagnetic character much more pronounced and, as a consequence, this material has an undesired M-shape magnetic entropy change curve. Thus, this work contributes by (i) improving the sol-gel method for producing manganites nanoestructured (ii) providing a broader and useful magnetic entropy change  and (iii) pointing out  the existence of a threshold size particle, below which the magnetocaloric features lack utility.
Further works to improve this effect is to (i) prepare advanced nanomaterials shell-protected, i.e., magnetic nanoparticles with magnetic shell and (ii) determine the threshold nanoparticle size to avoid superparamagnetic character.

\section{Acknowledgements}
We acknowledge FAPERJ, CAPES, CNPq and PROPPI-UFF for financial support.

Author contributions: R.J.C.-V., V.M.A., S.S.P. and D.L.R.: sample preparation, X-ray analysis; M.S.R. and T. C.-S.: Theoretical model; A.P.C.C. and D.L.R.: SEM measurements and analysis; A.A.C.: magnetic measurements. All authors discussed results and contributed to write the text of this paper.

\section{References}





\end{document}